\documentstyle[preprint,aps,epsf]{revtex}

\newcommand{\beq}{\begin{equation}}
\newcommand{\eeq}{\end{equation}}
\def\bea{\begin{eqnarray}}  \def\eea{\end{eqnarray}}

\begin{document}

\begin{center}
{\Large \bf Universality of the transverse momentum distributions in the
framework of percolation of strings} \\[8mm]
{\bf  J. Dias de Deus$^1$, 
E. G. Ferreiro$^2$, C. Pajares$^{2}$
and R. Ugoccioni$^{3}$}\par
{\it $^1$CENTRA, Instituto Superior T\'ecnico, 1049-001 Lisboa, Portugal}\par
{\it $^{2}$Departamento de
F\'{\i}sica de  Part\'{\i}culas and Instituto Galego de F\'{\i}sica de Altas Energ\'{\i}as,
Universidade de Santiago de
Compostela,\\ 15782--Santiago de Compostela, Spain}\par
{\it $^{3}$Dipartimento di 
Fisica Teorica and INFN-sezione di Torino, via P. Giuria 1,
I-10125 Torino, Italy}

\vskip 1.0truecm
{\Large {\bf Abstract}}
\end{center}
\vskip -0.5cm
In the framework of percolation of color sources, the transverse momentum
distribution in heavy ion and $p+p$ collisions at all centralities and energies
 are shown to follow a universal behaviour.
The width of the distribution depends on the fluctuations of the number of
color sources per cluster. At low densities, there are only independent single color sources,
no fluctuations occurs and the distribution is described by a single exponential.
At very high densities, only one cluster of many color sources appears and therefore
there are not fluctuations either 
and the hardness of the distribution is suppressed.
The Cronin effect in this framework is due to a maximum of the
fluctuations which decreases as the density increases. We obtain a good agreement
with experimental data including the low $p_T$ behaviour and the spectra for
different particles. We show that the transverse momentum and multiplicity distributions
are related to each other in a defined way. This point is satisfied by the experimental
data on $p+p$ collisions at different energies.



\newpage
\section{Introduction}
Many data have already been collected and analized during past few years at the 
Relativistic Heavy Ion Collider (RHIC) in order to obtain a complete understanding 
of the dense QCD matter which is created in high energy heavy ion collisions.

These data show that the inclusive high $p_T$ hadron production
in $Au+Au$ central collisions 
is strongly suppressed
\cite{1}-\cite{3}
compared to the scaling with the number of binary nucleon-nucleon collision, 
$N_{coll}$, expected on the basis of the factorization theorem for hard processes
in perturbative QCD(pQCD)
\cite{4}. 
The suppression is larger at 
forward than at mid rapidities \cite{5}.
No suppression was found in $d+Au$ collisions
\cite{6}-\cite{7}
at mid rapidity. Furthermore, 
the proton and antiproton yields become similar to the pion one at 
$p_T \approx 1-2\ GeV/c$ 
\cite{6}. 
The ratio between the yields from  
central and peripheral $Au+Au$ collisions is larger for protons 
than for pions at $p_T\approx 2-5\ GeV/c$ \cite{6}.
On the other hand, at low $p_T$, the yield for 
pions is larger than the one for kaons, and both are larger than 
the one for protons \cite{8}-\cite{9}. 
At very low $p_T$, 
the spectra for all species show a characteristic behaviour \cite{9}. 

The data also show 
disappearance of back to back jet-like hadron correlations in $Au+Au$ collisions, 
contrary to what is observed in $d+Au$ and $p+p$ collisions
\cite{10}-\cite{11}
and a peculiar behaviour
of the fluctuations in transverse momentum
\cite{12}-\cite{13},
with a maximum at a certain centrality given by a number of participants around 
$N_{part} \approx 150$.

All these data have originated a lot of  discussion 
and different 
explanations,
aimed
to discriminate which of 
these effects are caused by initial state 
interactions and which ones are a consequence of final state interactions.

Here we are going to concentrate on the explanations based on initial state
interactions.
Different mechanisms of initial state interactions
have been proposed, like saturation in high density QCD
through the Colour Glass Condensate (CGC) \cite{14}-\cite{16} or clustering of strings \cite{17}
in the string models.
Another possibility is the shadowing through pomeron interaction in the Dual Parton Model \cite{17b}.
All these mechanisms have in common the modification of a multiple scattering pattern --in the target
rest
frame-- or gluon interaction --in a fast moving frame--.

In the framework of the Colour Glass Condensate, the number of gluons 
in the hadron wave function 
can reach saturation when their momenta become smaller than a saturation scale 
$Q_s(x)\sim x^{-\lambda s}A^{1/3}$. Clustering and percolation 
of color sources, strings or partons,
can also be seen as an initial state interaction phenomenon. 
In the case of strings, these are formed by color fields stretched between partons of the projectile and
the target, located at the ends of the strings. These partons are embedded in the wave function of
the projectile and the target \cite{17c}.
In this case, above a critical density, a cluster of color sources --strings--
is formed through the whole collision area. 

In both approaches, if the number of partons exceeds a critical quantity,
they will begin to overlap in the transverse plane, and interact with each other, which prevents 
further growth of parton densities.
As the number of strings is mainly determined by the number of inelastic parton--parton collisions,
the density of strings is not a property of the isolated projectile. Nevertheless, since the density of
strings is connected with the density of partons and since the interaction and
percolation of strings takes place before the formation of final secondaries --before fragmentation--,
we call our effect {\it initial}. 


From this point of view,
the suppression of $p_T$ and the reduction of multiplicities
have the same origin 
\cite {18}-\cite {19}, namely, the overlapping of color sources. 
In both approaches, the transverse momentum
distributions satisfy a scaling law which is in agreement with the experimental data. 
In the CGC, the suppression was predicted to be stronger outside mid rapidity and
this point was later confirmed \cite {5}. 
Below, we will show that this is also the case 
in the framework of percolation of strings.
A more quantitative comparison of the 
results of both approaches shows also a remarkable agreement \cite{19b}.
The fact that the results of 
the string clustering approach, which is a soft, QCD inspired but model dependent description, 
coincide with the results of the CGC picture, which is a theory deduced from perturbative
QCD, induce us to think about the possibility of a smooth transition between soft and hard regimes
and about a {\it perturbative confinement} \cite{19c}.

All these features, together with a correct description of the transverse momentum 
fluctuations and of the low $p_T$ data, 
point out the existence of initial state interactions. 
On the other hand, 
the disappearance of back to back jet-like hadron correlations in $Au+Au$ collisions, 
contrary to what happens in $d+Au$ collisions, seems to point out a final state interaction
description such as jet-quenching \cite {20}  or interactions with partons and comovers \cite {21}.

The Cronin effect is crucial in order to discriminate which kind of interaction is working \cite{21b}. 
In this paper we will show that this effect is a low energy one, and it
is going to disappear at higher energies and/or densities. Similar results 
are obtained in the framework of the CGC 
\cite{22}.

Before starting with the description of our model, let us remember some general
features concerning the clustering.
On general grounds, in the clustering approach, one has to distinguish between 
two density regimes, high and low. 
Let us give a very simple example: the problem of throwing $N_S$ coins into $M$ boxes 
(one can read: production of $N_S$ partons or strings in the interaction area $\pi R^2$). 
The distribution $P(N)$ of $N$ coins in a box (one can read: N partons or strings in a cluster) 
can be studied and, in particular, the inverse normalized fluctuation $k$

\beq
	k=\frac{<N>^2}{<N^2>-<N>^2}
	\label{ec1}
\eeq
can be analytically calculated. At small density, $\eta=\frac{N_S}{M}<<1$, 
the coins are isolated and $k\rightarrow\infty$. At large density, $\eta>>1$, 
the coins are equally distributed in the boxes and, again $k\rightarrow\infty$. 
At some intermediate value $\eta_{min}$, there is a minimum. In the low density regime
$\eta<\eta_{min}$, $k$ decreases as $\eta$ increases, while in the high density
regime $\eta >\eta_{min}$, $k$ increases with $\eta$.

This behaviour for the fluctuations in the number of partons or strings 
in a cluster will be crucial in order to explain most of the data. 
In this paper we will explore these points in the framework of 
string percolation, using the strings as our basic objects. 
The use of strings could be considered as very model dependent. 
Nevertheless,
 the string structure can be derived directly from QCD under certain approximations
\cite{23}-\cite{25}. 
In fact, a string profile can be derived from the study of correlators of gluons fields, 
profile which is in agreement with the results obtained independently in lattice QCD \cite{25}.

In order to apply these ideas, we will use the following ingredients:
\noindent 
We need to know the number of strings $N_S$.
Up to RHIC energies, $N_S$ 
in the central rapidity region 
is approximately twice the number of collisions, $N_{coll}$.
However $N_S$ can be larger at RHIC and LHC energies. 
We compute $N_S$ using the Quark-Gluon String Model,
 equivalent to the Dual Parton Model \cite{26}-\cite{27}. 
Most of the reasonable string models 
\cite{28}-\cite{34} of heavy ions collisions obtained similar results for 
$N_S$. This fact gives us confidence in our values. Notice that sometimes, even in experimental 
analyses, $N_{coll}$ is obtained from the Glauber model without 
taking into account the energy-momentum conservation. At high energy this conservation reduces $N_{coll}$. 
In our computation of $N_S$ we use a Monte-Carlo code \cite{34}, based on the
Quark-Gluon String Model,
which takes into account 
energy-momentum conservation.

In order to obtain analytical 
formulae which can give an insight of 
the physical grounds we use soft $(exp(-ap^{2}_{T}))$ transverse momentum 
distribution \cite{35} and Poisson-like 
multiplicity distribution for the fragmentation of one string. 
We are aware that a string can also produce hard particles as in the Lund string model \cite{36}, 
and our simplification must be seen as a first approximation which clearly 
would fail at very high $p_T$. In spite of this, we will show how the clustering of 
strings gives rise to a universal behaviour of both the transverse momentum 
distribution and the multiplicity distribution, which are related to each other
through a gamma distribution that represents the cluster weight function. 
By the overlapping of soft strings, the soft spectrum gets a hard-like contribution. 
This behaviour must be seen as complementary to what happens in parton saturation in the CGC, 
where the clustering of 
gluons results in a softer spectrum. 
At RHIC and LHC energies, 
there is a large range of $p_T$ where both descriptions should coincide. 
Of course, we do not claim to give a description for the whole $p_T$ range at fixed energy.
Our explanation applies from low to
intermediate $p_T$ range. This range increases with the energy.
At very high $p_T$, a perturbative QCD description will be necessary.

The plan of the paper is as follows: first we 
describe our approach and 
we derive the transverse momentum and multiplicity distributions, in order 
to compare them with experimental data in the next chapter. The comparison includes
$p+p$ data and 
predictions for LHC energies. After, we discuss 
the effect of the clustering on the disappearance
of back to back jet-like hadron correlations. 
We will finish with some conclusions.

\section {Percolation of strings, transverse momentum and multiplicity
distributions}
Multiparticle production is currently described in terms of color strings 
stretched between the partons of the projectile and the target. 
These strings decay into new ones by sea $q-\overline{q}$ production, and subsequently 
hadronize to produce the observed hadrons. The color in these strings 
is confined to a small area in the transverse space, 
$\pi r^{2}_{0}$, with $r_{0}\simeq 0.2-0.25$ fm. 
This value is obtained in the vacuum correlator method and corresponds 
to the correlation length of the QCD vacuum. 
This value is in accordance with lattice results.

With increasing energy and/or atomic number of the colliding particles, 
the number of exchanged strings grows, and they start to overlap, forming clusters, 
very much like disks in the continuum two-dimensional percolation theory. 
At a certain critical density $\eta_{c}=1.18-1.5$ 
(depending on the type of employed
profile functions --homogeneous or Wood-Saxon--)
a macroscopical cluster appears, which marks the percolation phase transition.
For nuclear collisions, this density corresponds to $\eta= N_S \frac{S_1}{S_A}$
where $N_S$ is the total number of strings created in the collision, 
each one of an area $S_1=\pi r^{2}_{0}$ and $S_A$ corresponds to the 
nuclear overlapping area so it depends on the impact parameter $b$. 
For very central collisions $b=0$ and $S_A=\pi R^{2}_{A}$. 

The percolation theory governs the geometrical pattern of the string clustering. 
Its observable implications, however, require the introduction of some 
dynamics in order to describe the behaviour of the cluster formed by several overlapping strings. 
We assume that a cluster of $n$ strings behaves as a single string 
with an energy-momentum that corresponds to the sum of the
energy-momenta of the overlapping strings \cite{17},\cite{18}, and
with a higher 
color field, corresponding to the vectorial sum of the color charges of each individual 
$\stackrel{\rightarrow}{Q_1}$ string. The resulting color field covers the area $S_n$ of the cluster. 
As $ \stackrel{\rightarrow}{Q^{2}_{n}}=(\sum^{n}_{1}\stackrel{\rightarrow}{Q_{1}})^{2}$, 
and the individual string colors may be arbitrarily oriented,
the average $\stackrel{\rightarrow}{Q}_{1i}$  $\stackrel{\rightarrow}{Q}_{1j}$ 
is zero, so $\stackrel{\rightarrow}{Q^{2}_{n}}= n\stackrel{\rightarrow}{Q^{2}_{1}}$. 
$\stackrel{\rightarrow}{Q}_{n}$  depends also on the area $S_{1}$ 
of each individual string that comes into the cluster, 
as well as on the total area of the cluster $S_{n}$

\beq
Q_{n}=\sqrt{n\frac{S_n}{S_1}}\,Q_1\ .
\label{ec2}
\eeq

We take $S_1$ constant and equal to a disc of radius $r_0$.  $S_n$ corresponds to the 
total area occupied by $n$ discs, which of course can be different for different configurations 
even if the clusters have the same number of strings. One could make reasonable 
alternative assumptions about the interaction among the strings, as it was studied previously \cite{18}, 
but the comparison with the data on the dependence of the strength of the 
two-body \cite{38} and three-body Bose-Einstein correlations of the multiplicities \cite{39}, 
clearly favours (\ref{ec2}).

Notice that if the strings are just touching each other, $S_n = nS_1$ and $Q_n=nQ_1$, 
so the strings act independently to each other. On the contrary, if they fully overlap 
$S_n=S_1$ and $Q_n=\sqrt{n}Q_1$, so we obtain a reduction of the color charge.
Knowing the color charge $Q_n$, 
one can compute the multiplicity $\mu_n$ and the mean transverse momentum 
squared $<p^{2}_{T}>_n$ of the particles produced by a cluster, which are proportional
 to the color charge and color field \cite{18}-\cite{19}, respectively
\beq
\mu_{n}=\sqrt{\frac{nS_{n}}{S_{1}}}\ \mu_{1}\quad,\quad <p^{2}_{T}>_{n}=
\sqrt{\frac{nS_{1}}{S_{n}}}\ <p^{2}_{T}>_{1}
\label{ec3}
\eeq
where $\mu_{1}$ and $<p^{2}_{T}>_{1}$ are the mean multiplicity and 
mean $p^{2}_{T}$ of particles produced by a single string. We observe
\beq
	\mu_n<p^{2}_{T}>_n=n\mu_1<p^{2}_{T}>_1 \quad ,\quad \frac{\mu_n}{<p^{2}_{T}>}_n
=\frac{S_n}{S_1}\frac{\mu_1}{<p^{2}_{T}>}_1\ .
\label{ec4}	
\eeq

The first relation denotes that the product is an extensive quantity, 
while the second one indicates that each cluster satisfies a scaling law that it is 
nothing but the Gauss Theorem. From the Schwinger formula, one obtains 
$\mu_1=S_1 <p^2_T>_1$. From the experimental data we can fix
$\mu_1$ and $<p^2_T>_1$, obtaining $r_0\simeq 0.2-0.25$ fm in agreement with the
above mentioned QCD result.

Moreover, in the limit of high density $\eta$, one obtains 
\beq
<n\frac{S_{1}}{S_{n}}>=\frac{\eta}{1-\exp{(-\eta)}}\equiv\frac{1}{F\,(\eta)^{2}}
\label{ec5}
\eeq
and the eqs. (\ref{ec3}) transform into the analytical ones \cite{18}
\beq
\mu=N_{strings}\, F(\eta)\, \mu_{1}\quad,\quad <p^{2}_{T}>=\frac{1}{F(\eta)}\, <p^{2}_{T}>_{1}
\label{ec6}
\eeq
where $\mu$ and $<p^{2}_{T}>$ are the total multiplicity and mean 
transverse momentum and 
$N_{strings}$ is the total number of created strings 
in the considered rapidity range.

In the mid rapidity region, the number of strings $N_{strings}$ is proportional 
to the number of $A+A$ collisions, $N_{coll}\sim N^{4/3}_{A}$ \cite{40}-\cite{41}, 
being $N_A$ the number of wounded nucleons of one nucleus. 
In this case, the density of strings becomes $\eta=N_{strings}\frac{S_1}{S_A}\sim N^{2/3}_A$.
At high densities, one should consider the percolation limit -all the 
strings overlap into a single cluster that occupies the whole nuclear overlap are-. 
In this case, from eq. (\ref{ec3}) one obtains 
$$\mu_{A} = \sqrt{\frac{N_{strings}\cdot S_{A}}{S_1}} \mu_1,
\quad {\rm where} \ S_A= \pi R^2_A \propto N^{2/3}_{A} \ .$$
In other words, the multiplicity per participant becomes independent of $N_A$, i.e. {\it saturates}. 

Outside mid rapidity, $N_{strings}$ is proportional to the number of participants $N_A$
instead of to the number of collisions
$N^{4/3}_A$. Therefore, there is an additional suppression factor $N^{1/3}_{A}$ 
compared to central rapidity. This fact is at the origin of the larger suppression 
at $y=3$ of the observed $p_T$ distributions in $Au+Au$ and $d+Au$ collisions.

We use eq. (\ref{ec3}) to compute the multiplicities, using a Monte-Carlo code based 
on the Quark-Gluon String Model to generate the strings \cite{34}. Each string is produced 
at an identified impact parameter. From this, knowing the transverse area of each string, 
we identify all the clusters formed in each collisions and subsequently compute for each 
of them its multiplicity in units of $\mu_1$. The value of $\mu_1$ was fixed 
by the comparison of our results with WA 98 data for $Pb+Pb$ central collisions. 
Our results are in agreement with SPS and RHIC multiplicity data \cite{19}.
Using the first of the eqs. (\ref {ec6}) we obtain very similar results \cite{41}. 
We observe a weak dependence on $N_A$ of the rapidity density per participant at 
high centrality, 
i.e. saturation. On the other hand, in the fragmentation region, we expect the particle density per 
participant nucleon to be equal or even less than the nucleon-nucleon rapidity density \cite{41}. 
Both features, saturation and fragmentation scaling, are in agreement with experimental data.

Until now we have presented our results concerning mean values --mean multiplicity and
mean $p_T$--. In order to get distributions, we will develop the following strategy:
we will introduce the multiplicity and the transverse momentum distribution for the 
fragmentation of one string. We will weight them with the cluster function,
and then we will introduce the effect of the overlapping through $F(\eta)$. We will check 
that we obtain the relation (\ref{ec6}) for the mean values.

The multiplicity distribution in heavy ion collisions can be expressed
\cite{42}
as a superposition of
Poisson distributions with different mean multiplicities,
\beq
P(n)=\int dN\ W(N)\ P(N,n)
\label{ec7}
\eeq
The Poisson distribution
$P(N,n)=\frac{{\rm e}^{-N} N^n}{n!}$, $N=<n>$,
represents the cluster fragmentation, while the
weight factor $W(N)$ reflects the mean multiplicity distribution of the clusters,
related to the cluster size distribution and to the number of strings per cluster.
This quantity has contributions
due to both the nuclear structure and the parton distribution inside the
nucleon.

Concerning the transverse momentum distribution,
one needs the distribution $f(x,p_T)$ for each string or cluster, and the
mean squared transverse momentum distribution of the clusters, $W(x)$, 
which is also related to the cluster size distribution through the cluster tension.
For $f(x,p_T)$ we assume
the Schwinger formula, $f(x,p_T)=\exp(-p_T^2 x)$, used also
for the fragmentation of
a Lund string. In this formula $x$ is related
to the string tension, or equivalently to the mean transverse
mass
of the
string.
Assuming that a cluster behaves similarly to a single string
but with different string tension, that depends on the number of strings that
come into the cluster,
we can write for the total $p_T$ distribution
\beq
f(p_t)=\int W(x)\ f(x,p_T)\ .
\label{ec8}
\eeq
The weight function $W(x)$ 
obeys the gamma distribution
\beq
W(x)=\frac{\gamma}{\Gamma(k)} (\gamma x)^{k-1}\ \exp{(-\gamma x)}\ .
\label{ec9}
\eeq
The reason to chose a gamma distribution is the following:
in peripheral heavy ion collisions, the density of strings is small
and therefore there is no overlapping. The cluster size distribution in this
case is peaked at low values. As the centrality increases, the density of
strings also increases, so there is more and more overlapping among the
strings. The cluster size distribution is strongly modified, according to
the picture shown in Fig. 1, 
where we have plotted 
three cluster distributions that correspond to three increasing
centralities
of the collision.
Each curve in Fig. 1 can be compared to a gamma distribution, with different
$k$ values.
Moreover, the increase of centrality
can be seen as a transformation of the cluster size distribution of the type
\beq
P(N) \longrightarrow \frac{N P(N)}{<N>}
\longrightarrow .\ .\ .\
\longrightarrow \frac{N^k P(N)}{<N^k>}
\longrightarrow .\ .\ .\
\label{ec10}
\eeq
This kind of transformations were studied long ago by Jona-Lasinio in connection to
the renormalization group in probabilistic theory \cite{43}.
Actually, an increase of the centrality implies a transformation from clusters
with very few strings to another set of cluster with a higher number of strings, as can be seen 
in Fig. 1, and a
renormalization of the main variables of the clusters, i.e. mean transverse momentum and mean multiplicity,
induced by the higher color
of the new clusters. 
These transformations
have been used to study the probability distribution associated to rare events
\cite{44}-\cite{45}.

Introducing (\ref{ec9}) into (\ref{ec7}) and (\ref{ec8}) we obtain
\beq
\frac{\Gamma(n+k)}{\Gamma(n+1) \Gamma(k)} \frac{\gamma'^k}{(1+\gamma')^{n+k}}=
\int_0^\infty  dN
\frac{{\rm e}^{-N} N^n}{n!}
\frac{\gamma'}{\Gamma(k)} (\gamma' N)^{k-1} \exp{(- \gamma' N)}
\label{ec11}
\eeq
and
\beq
\frac{1}{(1+\frac{p_T^2}{\gamma})^k}=
\int_0^\infty  dx
\exp{(-p_T^2 x)}
\frac{\gamma}{\Gamma(k)} (\gamma x)^{k-1} \exp{(- \gamma x)}\ .
\label{ec12}
\eeq
The distribution obtained in (\ref{ec11}) is the well known negative binomial
distribution, whose
mean value and dispersion verify
\bea
<n>=<N>=\frac{k}{\gamma'}\ ,\ \
\frac{<N^2>-<N>^2}{<N>^2}=\frac{1}{k}\ ,\ \
\frac{<n^2>-<n>^2}{<n>^2}=\frac{1}{k}+\frac{1}{<n>}\ .
\label{ec13}
\eea
In the distribution (\ref{ec12}) the corresponding values are:
\bea
<x>=\frac{k}{\gamma}\ ,\ \ \ \ \ \
\frac{<x^2>-<x>^2}{<x>^2}=\frac{1}{k}\ .
\label{ec14}
\eea
The parameters $\gamma$ and $\gamma'$ are different,
since $<N> \ne <x>$,
while $k$ is the same in
both equations.

The equations (\ref{ec11}) and (\ref{ec12}) can be seen as a superposition of
chaotic sources -clusters- where
$\frac{1}{k}$ fixes the transverse momentum fluctuations.
At small density, $\eta << 1$, the strings are isolated and $k \rightarrow
\infty$. When the density increases, there will be some overlapping of strings forming
clusters, increasing the denominator of eq. (\ref{ec1}) and therefore decreasing $k$.
The minimum of $k$ will be reached when the fluctuations in the number of
strings per cluster reach its maximum. Above this point, increasing $\eta$,
these fluctuations decrease and $k$ increases.

The distribution
$W(x)$ satisfies the Koba-Nielsen-Olsen scaling, i. e. $<x> W(x)$ depends only
on $\frac{x}{<x>}$\footnote{However, it can depend on the energy through $k$.}.
This property \cite{46} stems from the type of the transformations (\ref{ec10}).

In order to take into account the effect of overlapping of strings on the multiplicity and
the mean transverse momentum, we need to included our factor $F(\eta)$
in the corresponding fragmentation functions, so the Poisson distribution
transforms into $P(N,n)=\frac{{\rm e}^{-N F(\eta)} (N F(\eta))^n}{n!}$ 
and the Schwinger formula transforms into
$f(x,m_T)=\exp(-m_T^2 x F(\eta))$. 
Indeed, what is happening is that
the invariance of the weight function
under the transformation $x \rightarrow \lambda x$ and
$\gamma \rightarrow \gamma/\lambda$ (or equivalently $N \rightarrow \lambda N$
and $\gamma' \rightarrow \gamma'/\lambda$), where $\lambda=F(\eta)$,
leads to the changes $<p_T^2> \rightarrow
<p_T^2>/ \lambda$ and $<n> \rightarrow \lambda <n>$ in the transverse mass
and multiplicity distributions respectively.

The multiplicity distribution becomes then the universal function
\beq
\frac{\Gamma(n+k)}{\Gamma(n+1) \Gamma(k)} \frac{(\gamma'/F(\eta))^k}{(1+
(\gamma'/F(\eta)))^{n+k}}=\frac{\Gamma(n+k)}{\Gamma(n+1) \Gamma(k)}
\frac{(k/(<n>_1 F(\eta)))^k}{(1+
((k/(<n>_1 F(\eta))))^{n+k}}
\label{ec15}
\eeq
and the transverse momentum  distribution behaves as
\beq
f(p_T,y) = 
\frac{dN}{dy} \frac{k-1}{\gamma/F(\eta)}
\frac{1}{(1 +\frac{p_T^2}{\gamma/F(\eta)})^k}=
\frac{dN}{dy} \frac{k-1}{k <p_T^2>_1} F(\eta) 
\frac{1}{(1 +\frac{F(\eta)\ p_T^2}{k\ <p_T^2>_1})^k}\ .
\label{ec16}
\eeq
The above equation has been normalized to
\beq
f(p_T,y) = \frac{dN}{dp_T^2 dy}\ .
\label{ec17}
\eeq

In the new distributions, $<p_T^2>=<p_T^2>_{old}/F(\eta)$ and $<n>=F(\eta)
<n>_{old}$, 
compared to eqs. (\ref{ec11}) and (\ref{ec12}). This agrees with our 
result from eqs. (\ref{ec6}).

We have, in particular
\beq
<n> = F(\eta) N_s <n>_1= F(\eta) \eta (\frac{R_A}{r_0})^2 <n>_1, \ \ \ \ 
<p^2_T>=\frac{k}{k-2}\frac{<p^2_T>_1}{F(\eta)}
\label{ec18}
\eeq
and the $p_T^2$ dispersion
\beq
Dp^2_T=<p^2_T>\sqrt{\frac{k-1}{k-3}}.
\label{ec19}
\eeq

The eqs. (\ref{ec15}) and (\ref{ec16}) summarize our main results. The multiplicity 
and transverse momentum distributions for any type of
collision and degree of centrality are universal functions which depend
only on one parameter, $<n_1>$ and $<p^{2}_{T}>_1$ respectively. 
These parameters are related to $\gamma'$ and $\gamma$ through eqs. (\ref{ec13}) and (\ref{ec14}).
Because of this, our parameters $\gamma$ and $\gamma'$ in the weight function 
are different.
The additional parameter $k$ depends on $\eta$ in the way that has been
pointed out before. It is related to the fluctuations in the number of strings
per cluster by eq. (\ref{ec1}) and plays an important role in the behaviour of the dependence
of the transverse momentum fluctuations on the number of participants \cite{47}-\cite{48}.
Notice that both distributions are obtained from the same kernel, 
the gamma distribution, with the same parameter $k$. This fact implies that they are related 
to each other.
In particular, the suppression of high $p_T$ production
in (\ref{ec16}) is controlled by $k$ which also provides us the width
of the multiplicity distribution at high multiplicities.
Due to this, we predict that in $p+p$ collisions at LHC energies, where the string
densities will be very high, the $k$ value will be higher and
in consequence 
both distributions, the multiplicity and transverse momentum one, 
will be narrower.

The relation between both distributions can be checked directly with the existing
experimental data from $p+p$ collisions. In fact, fixing the $k$-values from fits to the
multiplicity distributions, we have fitted the transverse momentum distributions,
with the power law $A(1+bp^{2}_{T})^{-k}$
obtaining an overall 
agreement with experimental data at $\sqrt{s}=23,200,630$ and 1800 GeV.
The detailed comparison is done in reference \cite{49}.

The values $<n>_1$ and $<p^2_T>_1$ stand for the average over all particle. 
If we are interested on the study
of a particular particle specie $i$, 
we shall write
$<n>_{1i}$ and $<p^2_T>_{1i}$.

At $\eta\rightarrow\infty$, $k\rightarrow\infty$ and the distribution (\ref{ec16})
becomes
$\exp(-F(\eta)p^{2}_{T}/<p^{2}_{T}>_{1i})$ very similar to the behaviour at $\eta\rightarrow 0,$\,\  
$\exp(-p^{2}_{T}/<p^{2}_{T}>_{1i})$.

From (\ref{ec16}) one can calculate
\beq
	\frac{d\ln f}{d\ln p_T}=
	\frac{-2F(\eta)}{(1+\frac{F(\eta)}{k}\frac{p^{2}_{T}}{<p^{2}_{T}>_{1i}})}
	\frac{p^2_T}{<p^2_T>_{1i}}\ .
	\label{ec20}
\eeq
As $p^{2}_{T}\rightarrow0$, this reduces to
\beq
	-2F(\eta) p^{2}_{T}/<p^{2}_{T}>_{1i}\ ,
	\label{ec21}
\eeq
while for large $p_T$, it becomes $-2k$.

We use a gamma distribution for the clusters, 
which 
is reasonable due to the above mentioned arguments.
It may be possible to use other distributions
which have the same properties concerning the dependence of their
mean value and dispersion on the centrality. The resulting multiplicity and transverse
momentum distributions in that case would be very similar to (\ref{ec15}) and (\ref{ec16}).
Nevertheless, we have additional confidence in our assumption, since the resulting multiplicity and
transverse
momentum distributions have a reasonable agreement with experimental data. In fact, the 
negative binomial distribution is in good agreement with the multiplicity distribution 
for $p+p$ collisions in a broad range of energies, 
and it can also describe the $h+A$ and $A+A$
multiplicity distributions.

The power-like behaviour $(p^{2}_{T})^{-k}$ found for the transverse momentum
distribution, with an exponent related to some intrinsic fluctuations,
 is 
common to many apparently different systems, as sociological, biological or
informatics ones \cite{50}-\cite{52}. Distributions that describe
the growth of the wealth of people living in stable economical systems,
the distribution of the citations of the
scientific works, or other complex networks where the probability $P(m)$ of having
a given node with $m$ links is described by the free scale power law $P(m)\sim(m)^{-k}$ with
$k$ close to 3 obey the same behaviour. Also, it has been shown \cite{52}-\cite{53} that maximization of 
the
non extensive information Tsallis entropy leads to the distribution (\ref{ec16}).

This universal behaviour indicates the importance of the common features
present in those phenomena, namely, the cluster structure and the
fluctuations in the number of objects per cluster.

\section {Density dependence of the mean transverse momentum and $k$}

A more specific  test of these ideas can be made in a straightforward manner from eqs. (\ref{ec18})
and (\ref{ec19}) and eq. (\ref{ec4}): there is a universal relation between 
$\left[\sqrt{\frac{<p^2_T>}{<p^2_T>_1}}\right]_i$ and $\sqrt{ \frac{1}{N^{2/3}_{A}} \frac{dn}{dy}}$,
\beq
\left[\sqrt{\frac{<p^2_T>}{<p^2_T>_1}}\right]_i=
\left(\frac{r_0}{R_1} \frac{1}{<n>_1}\right)
\sqrt{\frac{k}{k-2}}
\frac{1}{F(\eta)\sqrt{\eta}}
\sqrt{ \frac{1}{N^{2/3}_{A}} \frac{dn}{dy}}\ .
\label{ec22}
\eeq

A similar relation was suggested in the framework of the CGC model
\cite{54}, and a quantitative discussion was presented in \cite{55}. One should notice that (\ref{ec22})
is fairly independent of the kind of particle $i$ ($\pi$, k, p), 
independent of energy (from 200 GeV to 1800 GeV), and independent of the interacting nuclei ($Au+Au$ at
RHIC, $p+p$ at Tevatron).

From comparison with data, it can be 
checked that $k(\eta)$ is a function with the properties enunciated in the
introduction,
\beq
	k(\eta)=\frac{2\phi^2(\eta)}{\phi^2(\eta)-1}
	\label{ec29}
\eeq
 with
\beq
	\phi(\eta)=\sqrt{F(\eta)}+F(\eta)\sqrt{\eta}.
	\label{ec30}
\eeq

The function $k(\eta)$ has the qualitative behaviour that
we expect, namely, it has a minimum $k_{min} \simeq 3$ at $\eta_{min} = 2$ and
goes to infinite as $\eta \rightarrow 0$ and $\eta \rightarrow \infty$. 
This means that in such limits only one kind of cluster is produced.
However, the minimum value of $k$ is
reached at a slightly high value of $\eta$. There are reasons to
think about a modification of (\ref{ec30}), keeping the right limits for $k$ at
$\eta \rightarrow 0$ and $\eta \rightarrow \infty$, 
but introducing a shift in $\eta$. 
In fact, $F(\eta) \sqrt{\eta}=1-e^{-\eta}$ 
is the average fraction of the total area occupied by clusters at density $\eta$
when the area is homogeneous. Actually, in a heavy ion collisions, the surface
is not homogeneous when nuclei profile functions of the Wood-Saxons
type are taken into account
\cite{56}. 
This increases the average fraction of occupied area, 
resulting in a faster decrease of the exponential. Due to this, we use the
function given by eqs. (\ref{ec29}) and (\ref{ec30}) for $k$ but introducing
the shift $\eta\rightarrow b\eta$, 
in such a way that the minimum value of $k$ is reached 
at $\eta_{min}=0.6\ (b=3.3)$. In
Fig. 2 we present the functions before and after the change. Notice that in
both cases the minimum value is $k\simeq3$, where the
variance of the $p^{2}_{T}$ distribution diverges.

The main dependence of $k$ on the energy comes from the dependence on $\eta$.
In $A+A$ collisions, as the energy increases, the density of strings
increases and $k$ increases. However, there is an additional dependence
on the energy, working on the opposite way, which up to now has not been
taken into account. 
It is well known that even in hadron-hadron collisions 
at $\sqrt{s}\simeq30\ GeV$, there is enough energy to produce
hard scatterings. This modifies 
the single exponential
in the transverse momentum distribution. As far as the number of strings
is very low there is no possibility of overlapping. In perturbative QCD,
the study of the production of $n$ gluons leads to a distribution
whose width is controlled by $k\mu$, being 
$\mu=1/(1-\gamma)$,
$\gamma=\sqrt{\frac{6 \alpha_s}{\pi}}-\frac{11}{8}(1+\frac{2 n_f}{27})\frac{\alpha_s}{\pi}$.
As the 
energy increases, the QCD coupling constant $\alpha_s\rightarrow0$ and $\mu\rightarrow1$, in such
a way that asymptotically only the parameter $k$ remains.
As $k\mu$ decreases with the energy the distribution becomes broader.

To take into account this energy dependence we will use an effective
$k$ given by
\beq
	k_{eff}=k\ [a\ (\mu-1)+1]
	\label{ec31}
\eeq
with
$\mu$ as defined above. We use $\alpha_s(M_Z)=0.122$, so $\alpha_s (\sqrt{s}=19.4 {\rm GeV}) \approx 0.165$,
$\alpha_s (\sqrt{s}=200 {\rm GeV}) \approx 0.108$ and $\alpha_s (\sqrt{s}=1800 {\rm GeV}) \approx
0.081$.
The above expression satisfies $k_{eff}=k$ for $s\rightarrow\infty$. 
The values for $\alpha_s$ are lower that the ones used in other approaches for hadronic interactions,
especially at SPS energies --at low energy, threshold effects can be not negligible--. In other to take into
account these uncertainties, we have introduce the phenomenological expression (\ref{ec31}). 
In this way, the dependence of $k$ on $\alpha_s$ is modified through the factor $a$.
The value of
$a$, $a=0.3$, was determined by the comparison of the $p_T$ distributions for
$Pb+Pb$ collisions at SPS energies and 
$Au+Au$ collisions at $\sqrt{s}=200$ GeV with the experimental data.
From now on, in order to compare with experimental data, we will use
$k_{eff}$ from eq. (\ref{ec31}).
Through this equation, $k$ --solid line in Fig. 2-- is multiplied by a factor that depends on the energy. 
This factor is around 1.28 at SPS energies, 1.19 at RHIC energies and 1.13 at LHC energies (5500 GeV).  
Note
 that the dependece of $k$ on $\eta$ and on the energy is
model dependent, so it may suffer changes, specially at 
low energy where the threshold effects can be not negligible.
The fact that we want to stress here is the existence of a
minimum on the behaviour of $k$ versus $\eta$, related to a maximun
of fluctuations, and the limits of $k$ at low and high density, where $k\rightarrow\infty$.

\section {Comparison with the experimental data}
As we said before, the number of strings is computed using
a Monte-Carlo code \cite{34} based on the quark gluon string model. 
Knowing $N_S$ and $r_0 = 0.25$ fm, we can compute $\eta$ for each
type of collision. From eqs. (\ref{ec29})-(\ref{ec31}) 
we obtain $k$. Finally,
the values of $<p^{2}_{T}>_{1p}=0.30$, $<p^{2}_{T}>_{1k}=0.14$, $<p^{2}_{T}>_{1\pi}=0.06$
are normalized to reproduce the dependence of $<p^{2}_{T}>$ on 
centrality for protons, kaons and pions.
Note that the different $<p^{2}_{T}>_{1}$s correspond to different $\gamma$s in the original gamma
distribution

With these values, we have entirely determined eq. (\ref{ec16}) for all type of collision
energies and rapidities.

At low $p_T$, the behaviour given by eq. (\ref{ec21}) is clearly confirmed by PHOBOS data,
namely the derivative of $\log f(p_T)$ with $\log p_T$ vanishes as $p_T\rightarrow0$. 
As $<p^{2}_{T}>_{1p}$ $>$ $<p^{2}_{T}>_{1k}$ $>$ $<p^{2}_{T}>_{1\pi}$,
the absolute value is larger
for pions than for kaons and than for protons. At higher $p_T$, the distributions
become similar. This is shown in Fig. 3, where we present our results
for $Au+Au$ central collisions at $\sqrt{s}=200$ GeV, $\eta=2.5$, $k_{eff}=4$, 
together with
the PHOBOS data at low $p_T$
and PHENIX data at higher
$p_T$.
Notice that in the figure there are data from two different collaborations
with different normalizations. Probably, this is at the origin of the
minor differences between data and our results. (We do not fit the PHOBOS data,
but just apply equation (\ref{ec16}) fixing the normalization $\frac{dn}{dy}$
to reproduce the point at 
$p_T= 2$ GeV).

Now, let us discuss the interplay between low and high $p_T$.

One defines the ratio $R_{CP} (p_T)$ between central and peripheral collisions as
\beq
	R_{CP}(p_T)= \frac{f'(p_T, y=0)/N'_{coll}}{f(p_T, y=0)/N_{coll}}
	\label{ec32}
\eeq
where the distribution in the numerator corresponds to higher densities,
$\eta'>\eta$. The division by $N_{coll}$ essentially eliminates $N_S$ from
$\frac{dn}{dy}$ (this is true only at mid rapidity) and from (\ref{ec16}) and (\ref{ec6}) 
we obtain
\beq
	R_{CP}(p_T)=\frac{\left(\frac{k'-1}{k'}\right)}{\left(\frac{k-1}{k}\right)}
	\left(\frac{F(\eta')}{F(\eta)}\right)^2 
	\frac{\left(1+\frac{F(\eta)}{k}\frac{p^{2}_{T}}{^<p^{2}_{T}>_{1i}}\right)^k}
	{\left(1+\frac{F(\eta')}{k'} \frac{p^{2}_{T}}{<p^{2}_{T}>_{1i}}\right)^{k'}}\ .
	\label{ec33}
\eeq
In the $p_T\rightarrow0$ limit, 
taking into account that $\frac{2}{3}\leq\frac{k-1}{k}<1$
and that $F(\eta')< F(\eta)$, we obtain
\beq
	R_{CP}(0)\simeq \left(\frac{F(\eta')}{F(\eta)}\right)^2<1\ ,
	\label{ec34}
\eeq
approximately independent of $k$ and $k'$. As $\eta'/\eta$ increases,
the ratio $R_{CP}$ decreases, in agreement with experimental data.

As $p_T$ increases, we have
\beq
	R_{CP}(p_T)\sim 
	\frac{1+F(\eta)p^{2}_{T}/<p^{2}_{T}>_{1i}}
	{1+F(\eta')p^{2}_{T}/<p^{2}_{T}>_{1i}}\ ,
	\label{ec35}
\eeq
and $R_{CP}$ increases with $p_T$ (again, $F(\eta)>F(\eta'))$.

At large $p_T$,
\beq
	R_{CP}(p_T)\sim \frac{F(\eta)}{F(\eta')}
	\frac {k'}{k} {p^{2}_{T}}^{k-k'}\ ,
	\label{ec36}
\eeq
which means that if we are in the low density (low energy,
low $N_A$) branch of the $k(\eta)$ curve, see Fig. 2, $k>k'$
and $R_{CP}(p_T)>1)$ (Cronin effect). As $\eta'/\eta$ increases
the ratio $R_{CP}(p_T\rightarrow\infty)$ increases (it must have
a limit due to phase space limitations).

As we increase the energy of the nucleus-nucleus collision,
the energy density increases and one has to observe the high density
branch of the $k(\eta)$ curve. There, for $\eta'>\eta$, $k'>k$ and
suppression on $p_T$ occurs. The Cronin effect will disappear
at high energies and/or densities.

In the forward rapidity region, the division by $N_{coll}$ in (\ref{ec32}) does not 
cancel $N_S$ from $\frac{dn}{dy}$, since in this rapidity region $N_S$ is proportional to
$N_A$ instead of $N_{coll}$. Therefore, an additional factor $\frac{N'_A/N'_{coll}}{N_A/N_{coll}}$
appears now in $R_{CP} (p_T)$. As $N'_{coll}-N'_A$ is much larger than $N_{coll}-N_A$,
$R_{CP} (p_T,y=3) < R_{CP} (p_T,y=0)$, thus a further suppression occurs, in agreement with
experimental data \cite{56b},\cite{56c}.
In order to quantify this suppression, let us consider central and peripheral 
$d+Au$ collisions at mid rapidity and at $y=3.2$.
From the BRAHMS Collaboration data \cite{56b} at $p_T \simeq 2-3$ GeV/c,
$R_{CP} (p_T,y=3.2)/R_{CP} (p_T,y=0) \simeq 0.45/1.25 = 0.36$ for $0-20 \%$ and $60-80 \%$ 
degree of centrality.
For these centralities, the quoted values by the collaboration are 
$N'_A(d)=1.96$, $N'_{coll}=13.6$, $N_A(d)=1.39$ and $N_{coll}=3.3$. The 
resulting ratio $(1.96/13.6)/(1.39/3.3)=0.35$ is in perfect agreement with data.

The ratio between $R_{CP}(p_T)$ for two different particles,
for instance $p$ and $\pi$, becomes, at large $p_T$,
\beq
	\frac{R^P_{CP}(p_T)}{R^{\pi}_{CP}(p_T)}\simeq
	\left(\frac{<p^2_T>_{1P}}{<p^2_T>_{1\pi}}\right)^{k'-k}\ .
	\label{ec37}
\eeq

For $Au+Au$ central and peripheral collisions at $\sqrt{s}=200$ GeV, we have $k'=4$
and $k=3.6$ respectively. Therefore, the ratio is close to 2, in good 
agreement with experimental data.

In Fig. 4, we compare our results for $\pi^0$ production in central and 
peripheral $Au+Au$  
and $p+p$ collisions at $\sqrt{s}=200$ GeV, 
together with the experimental data. In Fig. 5, 
the nuclear modification factor 
for central and
peripheral $Au+Au$ collisions at mid rapidity are shown. In Fig. 6, 
we show the comparison
of our results for $d+Au$ collisions at $\sqrt{s}=200$ GeV and 
at midrapidity, and in Fig. 7 we show the corresponding
nuclear modification factor.
In Fig. 8 we compare
our results with experimental data 
\cite{56d}\footnote{We are aware that there is a big uncertainty concerning SPS data (see Ref. \cite{56e}
for more details). Nevertheless, we use these data since they have been obtained by the same 
experimental collaboration as the ones we have
used to fix the value of $\mu_1$ in our model. With this value we can reproduce also multiplicities at SPS
and RHIC energy. These considerations let us to have a consistent description for $p_T$ and multiplicity
distributions.}
for central and peripheral $Pb+Pb$ collisions
at SPS energies. 
We observe a reasonable agreement in all cases. 
Notice that the trend of data can be understood qualitatively from our 
eq. (\ref{ec16}) and the ratios $R_{CP}(p_T)$ or the nuclear modification factor
depend on the difference $k'(\eta')-k(\eta)$
(at large $p_T$) and on the ratio $\frac{F(\eta')}{F(\eta)}$ (at low
$p_T$), and not on the absolute values of $k'$, $k$, $F(\eta')$ and $F(\eta)$,
in such a way that the uncertainties in the computation of $\eta'$ and $\eta$
due to the values of $r_0$ and $N_S$ are essentially canceled. All
the explanation stems from the facts that $F(\eta')<F(\eta)$ for
$\eta'>\eta$ and $k'>k$ for large densities and $k'<k$ for
small densities. 
Finally, in Fig. 9, we show our prediction for LHC energies.

\section {Back jet-like hadron correlations}

One of the most interesting data from RHIC is the disappearance of back to back 
jet-like hadrons in $Au+Au$ collisions, contrary to what happens in $d+Au$ 
collisions. 
We think that the suppression of produced jets in central $A+A$ collision is 
indeed due to final state interaction -jet-quenching 
or interactions with partons and comovers-. Here, for completeness, 
we would like to 
evaluate
the interaction of a quark or gluon-jet with a final 
state formed by a cluster of strings corresponding to a given value of $\eta$.
We will follow the reference \cite{57}.
In Fig. 10 we show the formation of several strings in a 
nucleus-nucleus collision. At high density there will be 
overlapping of strings in the transverse space, forming clusters.
Above the percolation critical density, essentially
one cluster will be formed through the whole collision area.
This colorless cluster can act as a non thermalized quark
gluon plasma, where the color would no longer be confined to hadronic
or flux tube dimensions but to the whole available area of
the scattering. 

The string-like shape of the color fields
are oriented along the collision axis. 
Consider a 
hard parton-parton collision, which produces two
hardly scattered partons which are moving in the transverse plane
to the reaction axis and in opposite directions (Fig. 11). 
The interaction of these partons with QCD fields of the 
strings affects
the parton momentum distribution, which determines the distribution of 
secondary particles in the jet. In particular, bremsstrahlung gluon
radiation takes place in a tangential direction by the parton crossing the
gluon field of each string. 
This gluon radiation will eventually produce low $p_T$ particles. In other words, 
the jets are not going to disappear but to be degraded to lower $p_T$ particles.
This is at the origin of the difference found by STAR collaboration \cite{57a} between near and away side
$p_T$ distributions.

This interaction produces an asymmetry, the
projection of momentum of secondary particles in the jet along the
reaction axis (Fig. 10, axis z) becomes larger than in the transverse
direction (axis x).
Let us
considerer a quark $q_1$ passing through the flux strings
transverse to the string axis. We assume constant static uniform chromoelectric
field pointed along the reaction axis. Because in the static case
chromomagnetic fields
vanish, the force acting on the quark $q_1$ can be written as
$$\stackrel{\rightarrow}{F}=\frac{g}{2} \lambda^a_1 \stackrel{\rightarrow}{E^a}$$
where $g$ in the QCD coupling,
$\frac{g^2}{4\pi}=\alpha_s$ and
$\lambda^a$
are the Gell-Mann matrices.
First, we compute the change of the transverse momentum
of $q_1$ after crossing one string. 
Let axis $x$ and $y$ be the coordinate system in the
section where $q_1$ crosses the string, 
let axis $y$ be pointed along $q_1$'s direction of motion and
let $\xi$ be the distance between axis $y$ and line of
$q_1$ movement.
Then $q_1$ will cross the string
surface in points $(\xi,y_1(\xi))$ and $(\xi,y_2(\xi))$ and the momentum change will be
\beq
	\Delta\stackrel{\rightarrow}{p_1}(\xi)=
	\int\stackrel{\rightarrow}{F}(\xi,t)dt=
	\int^{y_2(\xi)}_{y_1(\xi)}\stackrel{\rightarrow}{F}(\xi,y) dy=
	\int^{y_2(\xi)}_{y_1(\xi)}g\frac{\lambda^a_1}{2} \stackrel{\rightarrow}{E^a}(\xi,y) dy\ .
	\label{ec38}
\eeq 

Doing the average over all $\xi$ we obtain
\beq
\Delta\stackrel{\rightarrow}{p_1}=
\frac {1}{\int^{r_0}_{-r_o} d\xi}
\int ^{r_0}_{-r_0} d\xi
\int^{y_2(\xi)}_{y_1(\xi)} g \frac {\lambda^a_1}{2}\stackrel{\rightarrow}{E^a}dy=
\frac {1} {2r_0}
\int_{s_1} ds\ g\frac{\lambda^a_1}{2}\stackrel{\rightarrow}{E^a}\ ,
\label{ec39}
\eeq
where
$S_1$ is the transverse string area. Applying the Gauss theorem we have
\beq
\Delta\stackrel{\rightarrow}{p_1}=
\frac{ \widehat{\stackrel{\rightarrow}{z}}}{2r_0}\frac{g\lambda^a_1}{2}
\int_{S_1} \stackrel{\rightarrow}{E} \stackrel{\rightarrow}{dS}=
\frac{ \widehat{\stackrel{\rightarrow}{z}}}{2r_0}\frac{g\lambda^a_1}{2}
\int_{V}div\stackrel{\rightarrow}{E^a}dV=
\frac{\widehat{\stackrel{\rightarrow}{z}}}{2r_0}\frac{g^2}{8} \lambda^a_1 \lambda^a_2
\label{ec40}
\eeq
where $\frac{g\lambda^a_2}{2}$ is the color charge of quark $q_2$ (string end) 
and $\widehat{z}$ is the unit
vector in the $z$ direction.
Doing the average over the color states of $q_1$ and $q_2$, we obtain
\beq
\Delta p_1=\frac {g^2}{8\pi_0} \frac {1}{2\sqrt{2}}\frac{16}{3}=
\frac{4\pi\alpha_S}{3\sqrt{2}r_0}\ .
\label{ec41}
\eeq
The average number of strings crossed by the quark is
\beq
N=2r_0 Lm
\label{ec42}
\eeq
where $L$ is distance traveled by the quark, $L\approx R_A$ and 
$m$ is the number
of strings per unit area
\beq
m=\frac{N_S}{\pi R^2_A}\ .
\label{ec43}
\eeq

In $A+A$ collisions at mid rapidity, $N_S \approx 2N^{4/3}_{A}$
at RHIC energies and $3N^{4/3}_{A}$ at LHC energies. In the
forward rapidity region $N_S$ will be $2N_A$ and $3N_A$ at RHIC and LHC energies
respectively.

The mean transverse momentum of quark $q_1$ after it crosses $N$ strings, would be
\beq
<\Delta p^2_{ZN}>=
<(\Delta p_{Z N-1}+ \Delta p_Z)^2>=
<\Delta p^2_{Z N-1}>+<\Delta p^2_Z> + 2<\Delta p_{Z N-1}\Delta p_Z>\ .
\label{ec44}
\eeq
As $\Delta p_{Z N-1}$ and $\Delta p_Z$ may be pointed in any direction along
the string independently, the last term vanishes and therefore
\beq
<\Delta p>_{tot} \simeq \sqrt{N} \Delta p_1\ .
\label{ec45}
\eeq
For $\alpha_S \simeq 0.3$ and $r_0 = 0.2-0.25$, we obtain from (\ref{ec41})
$\Delta p_1 \simeq 0.9 - 0.75$ GeV/c. For central $Au+Au$ collisions
at mid rapidity we obtain $\sqrt {N} \simeq 5.5 - 7$.
Then $<\Delta p>_{tot} \simeq 4.9 - 5.2$ GeV/c,
which is comparable with the $p_T$ triggered, therefore the back to
back jet-like correlations disappears. On the contrary,
for central $d+Au$ collisions, $\sqrt{N}\simeq 0.75 -0.85$ and
$<\Delta p>_{tot} \simeq 0.67-0.64$ GeV/c,
which is much smaller than the $p_T$ of the triggered jet, $p_T > 4$ GeV/c,
and the back to back jet-like correlation survives.
An intermediate situation is $Au+Au$ central collisions at forward
rapidity, where 
$\sqrt{N} \simeq 2.5 - 3.2$ and $<\Delta p>_{tot} \simeq 2.2-2.4$ GeV/c.
In this case, the back to back jet structure is only partially destroyed.
In the case of peripheral collisions, 
the suppression would be stronger when the
jet is perpendicular to reaction plane (the plane spanned by the
beam axis and the impact parameter, $b\neq 0$) than when the jet is in the
reaction plane, in agreement with experimental data. 
Notice that before we deduced stronger suppression at forward than at mid rapidities 
for the nuclear modification factor
but we predict
less suppression of the jet. In the first case, the origin
of the further suppression is the normalization of the clustering in the
initial state and in the second case, the lower interaction of the
quark jet with the final state is due to the smaller string densities present 
at forward
rapidities.
Finally, let us mention that, 
in the framework of the CGC, 
it has been also found
stronger depletion of the back to back correlations in
$p+A$ and $A+A$ collisions than in the case of $p+p$ collisions
\cite{57b}.

\section {Conclusions}
We have obtain a universal transverse momentum distribution 
that allows us to describe
the low $p_T$ shape of the different particle species and the suppression of the 
high $p_T$ 
yield compared to the scaling with
the number of collisions, $N_{coll}$,
expected on the basis of factorization theorem of QCD.

The shape of this distribution is determined essentially by two functions,
$F(\eta)$ and $k(\eta)$, which depend on the density of color sources $\eta$.
The first one stems from the dynamics of the color clusters, and it is
at the origin of the suppression of 
the multiplicity. It also controls the transverse momentum distribution
of the produced particles due to the fragmentation of the cluster.
The second one, $k(\eta)$, is related to the fluctuations in the number of
strings per cluster. At low density, $k$ decreases with the density
and, on the contrary, at high density it increases with the density.

The fact that for $p+Au$ and $d+Au$ central collisions the densities are 
not far from the minimum of $k$ explains the Cronin effect. 
It will disappear at higher energy or higher densities
because of the increase of $k$. In the forward rapidity region, the number 
of color sources scales with the number of participant nucleons, $N_A$, 
while in the central rapidity region it scales with the number of collisions
$N_{coll} \propto N_A^{4/3}$.
Therefore, comparing the $p_T$ yield
scaled by the number of collisions, 
in the forward rapidity region there will be 
a suppression due to the
additional factor $1/N^{1/3}_A$.

The multiplicity distributions are related to the transverse momentum ones 
through
the cluster size distribution, whose width is controlled by $k$. At high mean 
multiplicities the relative dispersion of the multiplicity distribution is
given by $1/k$, decreasing as the energy increases.

The disappearance of the back to back jet-like hadron correlations in $Au+Au$
collisions is a final state interaction due to the interaction of the
quark-jet with the cluster of strings formed in the initial state. This cluster
is less dense in the forward rapidity region, so we predict smaller suppression of
the correlation in this case.

We can conclude that the clustering of color sources 
provides us a framework suitable
to describe the main features of the transverse momentum and multiplicity
distributions at RHIC and SPS energies. This, together with the successful
explanation of the dependence of the fluctuations of transverse momentum on the
number of participants and the dependence on the multiplicity
of the strength of the two-body and three body correlations,
favors this approach.

We are aware of the limitations of our framework, coming mainly from the lack of a
direct derivation from QCD. Also, we do not included any hard component in the
single string. Therefore, at fixed energy and above some $p_{Tm}$ our picture
will fail. However, as the energy and the density grow, there are more overlapping
strings which extend the validity of the description to a higher $p_{Tm}$. Finally,
asymptotically, our picture is valid for all transverse momentum, recovering a single
exponential, $exp(-F(\eta) p^2_T/<p^2_T>_i)$.

Our approach has similarities with the CGC.
In both approaches there is clustering --of gluons in the CGC and of color sources in
our approach--.
In both approaches, the initial state interactions --gluon saturation
in the CGC or clustering of color sources in the percolation approach--
produce a suppression of the $p_T$ distributions.
In CGC the gluon field is renormalized as the color charge increases. In
our approach, we have to redefined our main variables for the cluster as the color
density increases. 
The suppression of the high $p_T$ yield and the saturation 
of the multiplicity per participant at high density are related to each other and 
they are a consequence
of the saturation of gluons at the scale given by the 
saturation momentum $Q_S$ in the CGC \cite{58}. 
The clustering of strings produces also the
suppression of $p_T$ and the independence of the multiplicity per participant on $N_A$. 
We
obtain, as in the CGC, larger suppression of the $p_T$ yield at forward rapidity, and also
we predict that the Cronin effect would disappear at higher energies. Both approaches
obtain a scaling for the transverse momentum distributions. We do an extension
of the soft spectra to the region of hard $p_T$ by means of clustering of strings.
The CGC does an extension of the hard region to the soft one by means of clustering
of gluons, therefore in a broad range of $p_T$ both approaches should coincide.

In QCD, the distribution for $n$ gluons is approximately given, in the modified
leading log approximation, by the generalized gamma function \cite{59}
\beq
\frac{\Gamma_n}{\Gamma}\sim \frac{1}{<n>}\left(\frac{n}{<n>}\right)^{k\mu-1}
exp\left(-D \frac{n}{<n>}\right)^\mu\ , 
\label{ec46}
\eeq
whose width becomes broader as the energy increases. According to our considerations
we expect a narrower width of this distribution above a certain density.
It would be welcome the proof of that.

Let us remark that other possibility of testing our approach is to study the
behaviour of the forward-backward correlations which are proportional to the
fluctuations in the number of independent clusters. At moderate density,
we expect a rise of these correlations as it is observed at $p+p$ and $h+A$ collisions.
On the contrary, at high density, these long range correlations should disappear as a
consequence of the formation of a large cluster of strings \cite{60}. 
Similar behaviour is expect in the CGC.

Finally, let us comment that our approach provide us with an initial state which
can affect the final state interactions, as it was shown above when we studied the
disappearance of back jet-like correlations.

This means that other phenomena as jet-quenching or interaction with comovers are not
necessarily contradictory to our approach. We only claim that at low and intermediate
$p_T$, $p_T\leq 4-5$ GeV/c, the percolation of color sources is able to give a 
simple and reasonable description of data.

We thank M. A. Braun, N. Armesto, F. del Moral, C. A. Salgado for discussions. This work
has been done under contracts FPA2002-01161 of CICyT of Spain, POCTI/36291/FIS/2000 of 
Portugal, Feder funds from EU and
PGIDIT03PXIC20612PN from Galicia.


\begin{figure}
\centering\leavevmode
\epsfxsize=6in\epsfysize=6in\epsffile{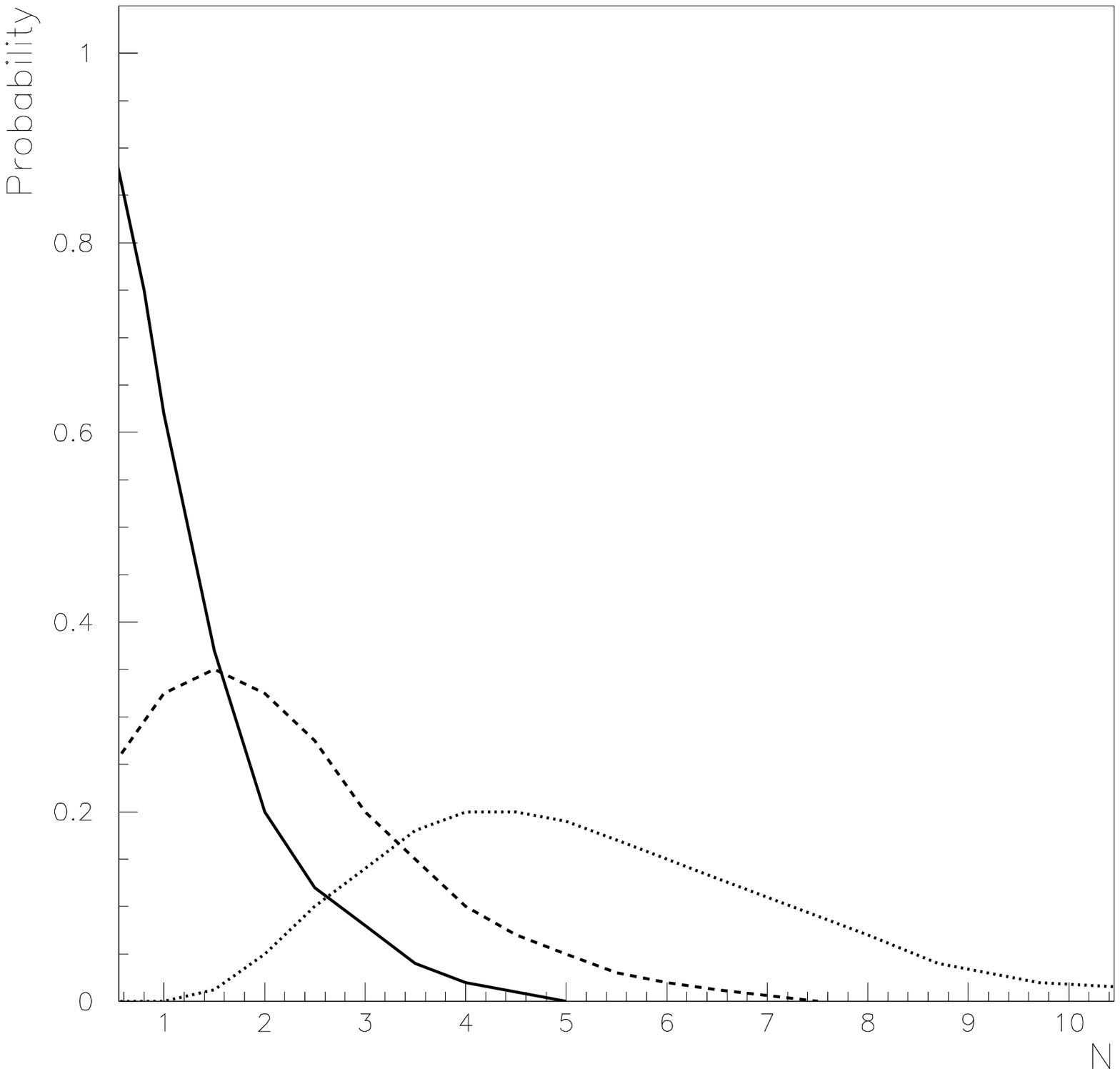}
\vskip 0.5cm
\caption{ Schematic
representation of the number of clusters as a function of the
number of strings of each cluster at three different centralities
(the continuous
line corresponds to the most peripheral one and the pointed line to the most
central one)}
\label{figure2}
\end{figure}

%
%
%
%

\begin{figure}
\centering\leavevmode
\epsfxsize=6in\epsfysize=6in\epsffile{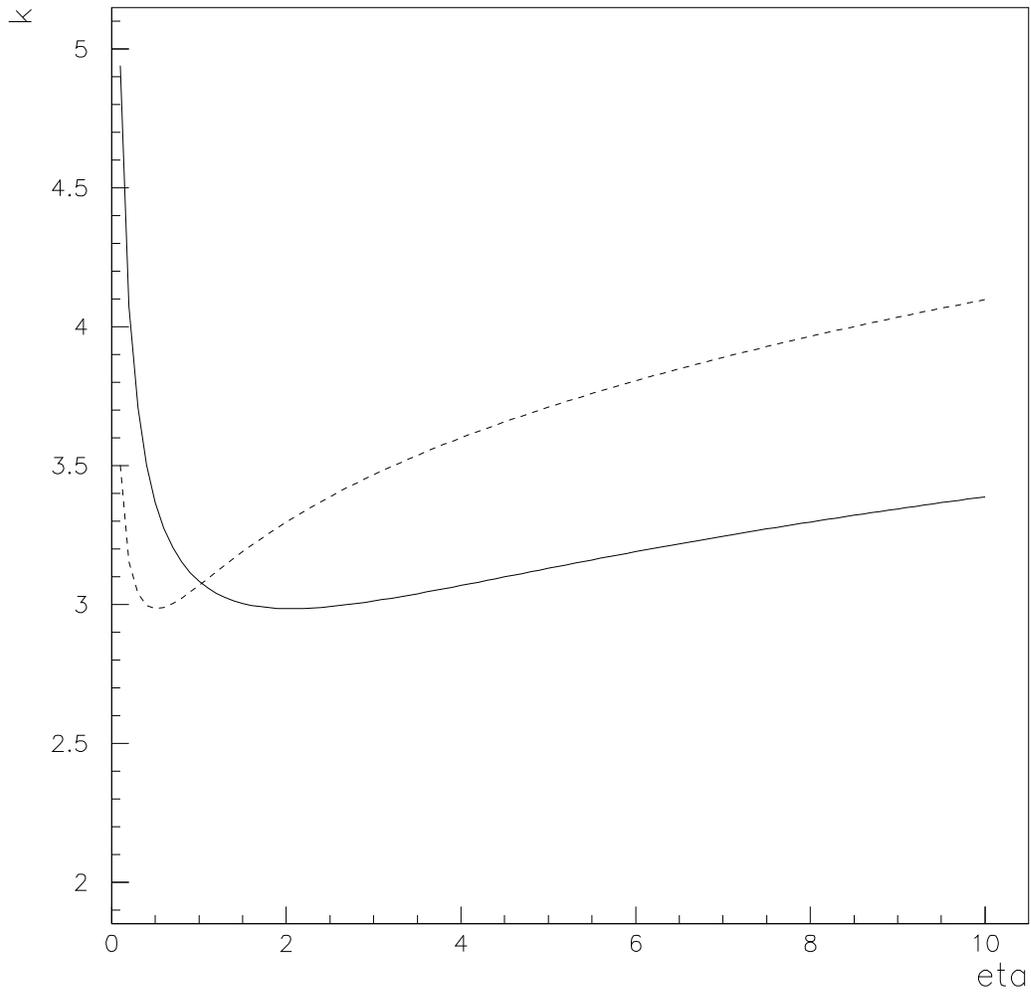}
\vskip 0.5cm
\caption{Dependence of $k$ on $\eta$, using eqs. (\ref{ec29}) and (\ref{ec30}) (dotted line)
and changing $\eta \rightarrow 3.3 \eta$ (solid line).}
\label{figure5}
\end{figure}

\begin{figure}
\centering\leavevmode
\epsfxsize=6in\epsfysize=6in\epsffile{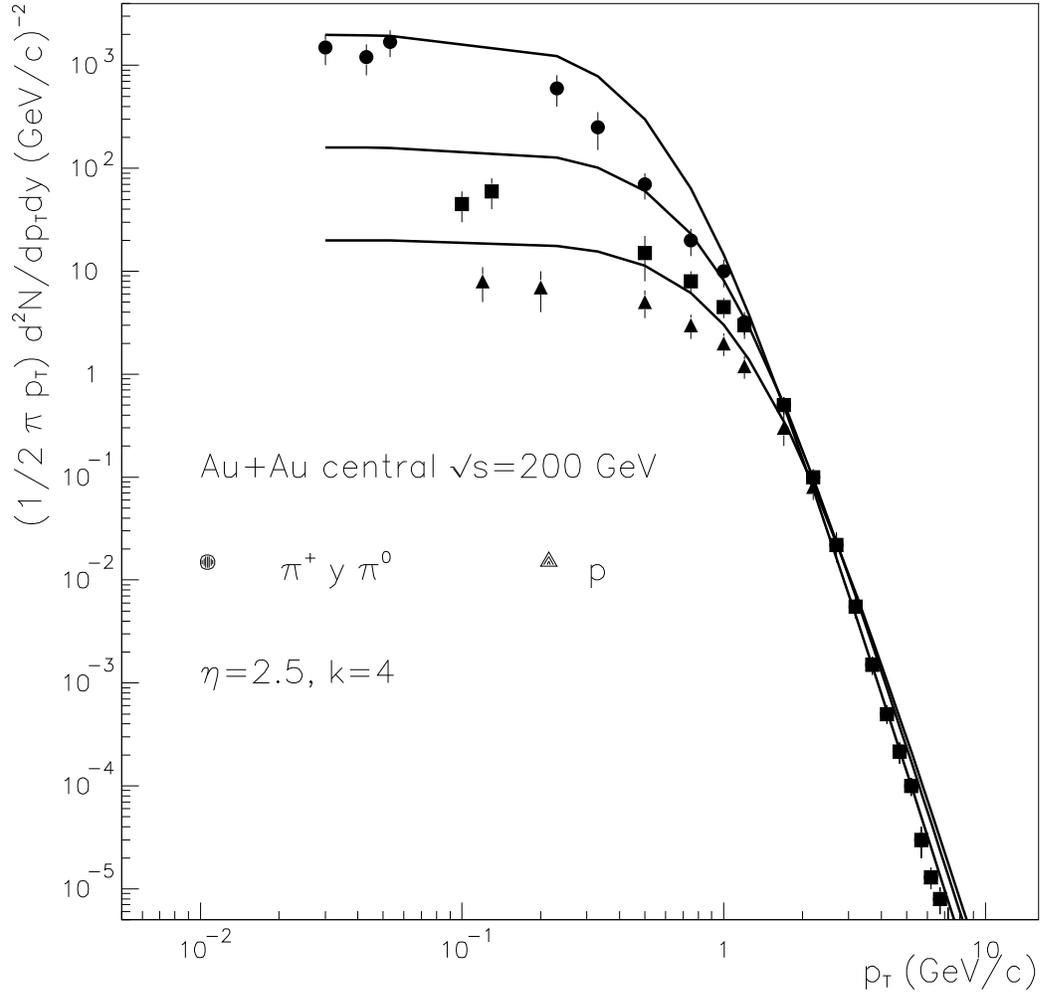}
\vskip 0.5cm
\caption{Experimental PHOBOS data on low $p_T$ distributions for pions, kaons and protons, together with
our results.}
\label{figure6}
\end{figure}

\begin{figure}
\centering\leavevmode
\epsfxsize=6in\epsfysize=6in\epsffile{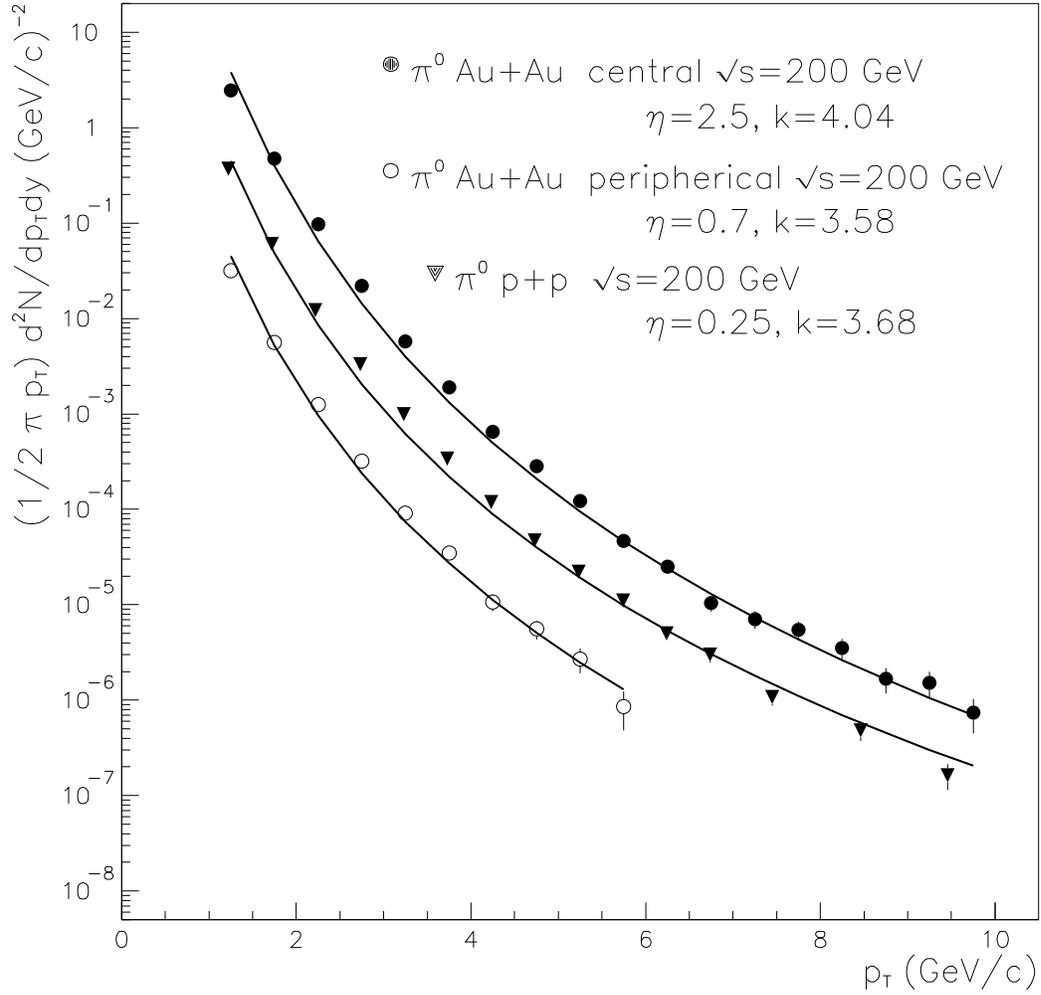}
\vskip 0.5cm
\caption{Comparison between our
results and experimental data from Au-Au central
and peripheral collisions and $p+p$ collisions at $\sqrt{s}=200$
GeV.}
\label{figure7}
\end{figure}

\begin{figure}
\centering\leavevmode
\epsfxsize=6in\epsfysize=6in\epsffile{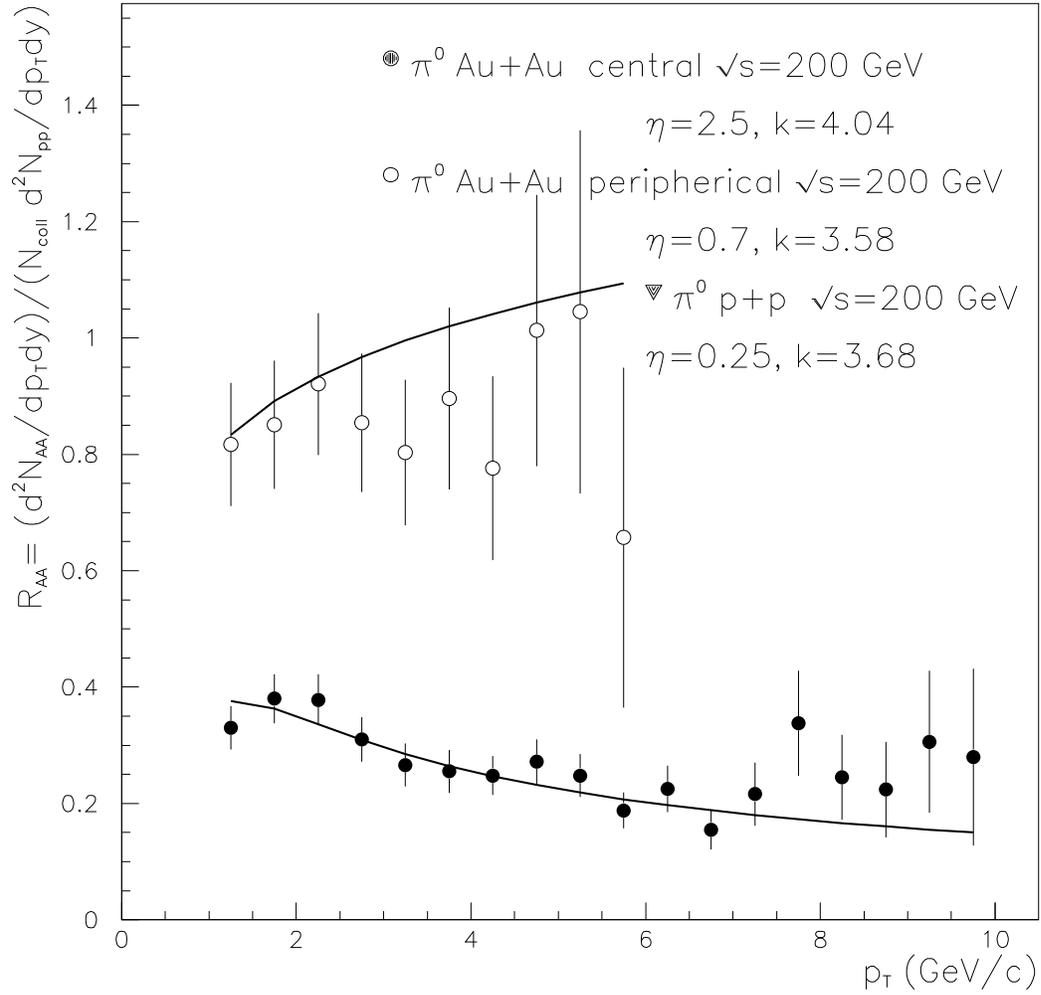}
\vskip 0.5cm
\caption{Comparison
between our results and experimental data 
on the nuclear modification factor 
from Au-Au central
and peripheral collisions at $\sqrt{s}=200$
GeV.}
\label{figure8}
\end{figure}

\begin{figure}
\centering\leavevmode
\epsfxsize=6in\epsfysize=6in\epsffile{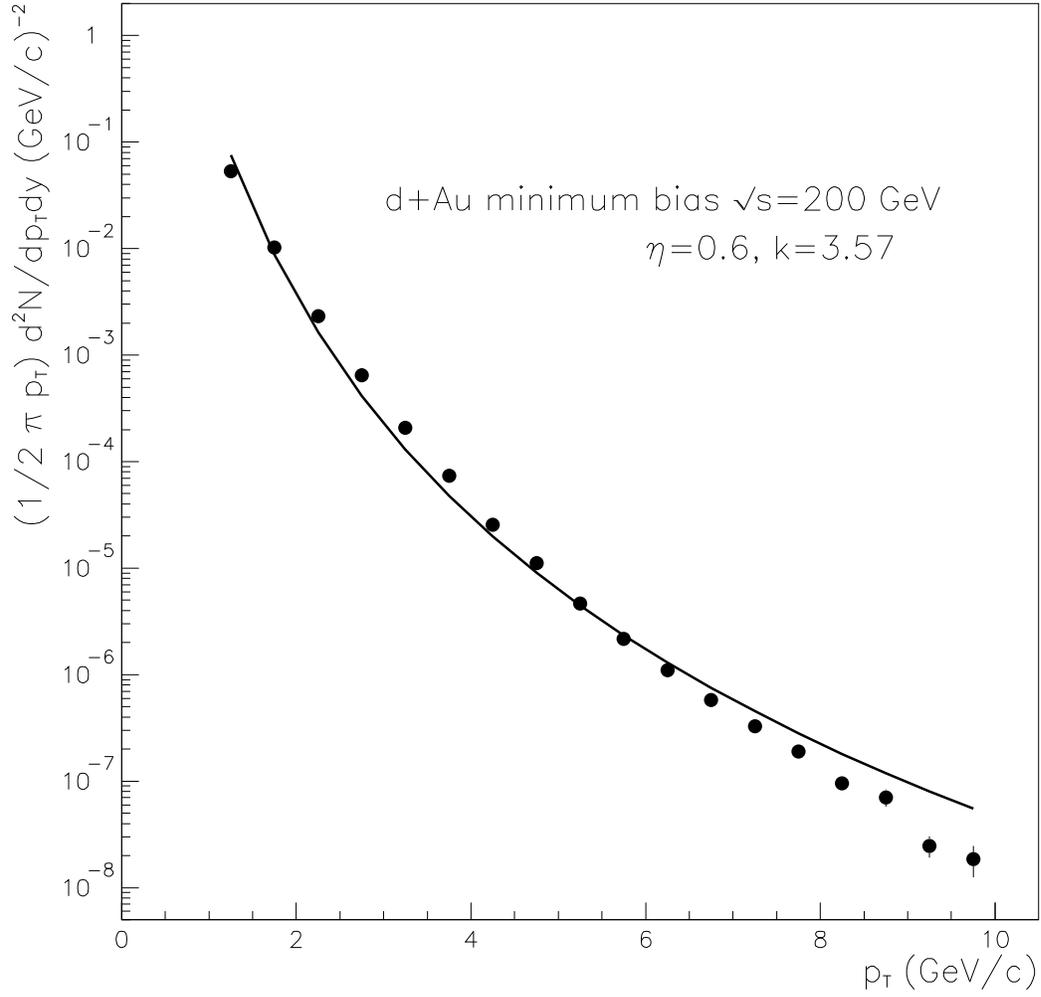}
\vskip 0.5cm
\caption{
Our results for d-Au 
minimum bias collisions at $\sqrt{s}=200$ and mid rapidity compared to experimental data.}
\label{figure9}
\end{figure}

\begin{figure}
\centering\leavevmode
\epsfxsize=6in\epsfysize=6in\epsffile{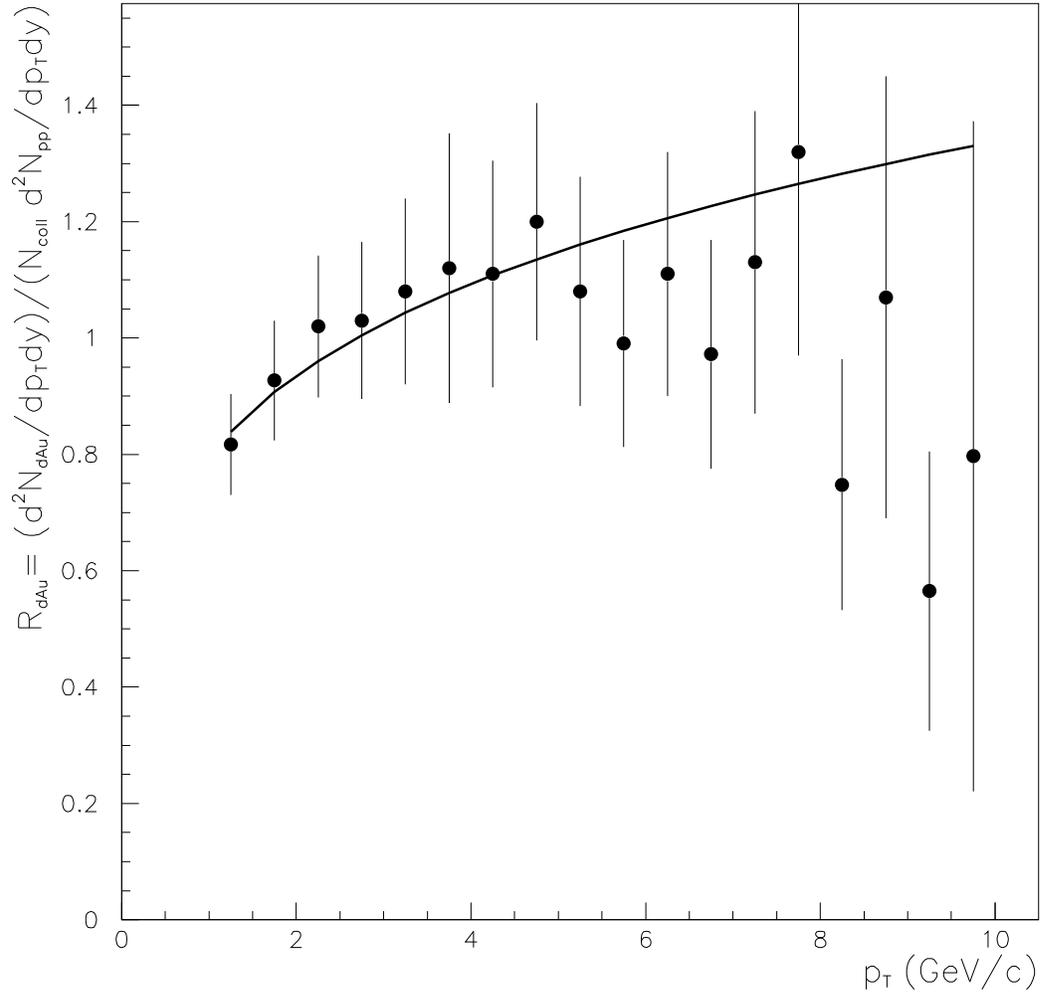}
\vskip 0.5cm
\caption{
Our results for d-Au 
minimum bias collisions at $\sqrt{s}=200$ and mid rapidity 
normalized to our results for p-p minimum bias collisions multiplied by the
number of collisions, compared to experimental data.}
\label{figure10}
\end{figure}

\begin{figure}
\centering\leavevmode
\epsfxsize=6in\epsfysize=6in\epsffile{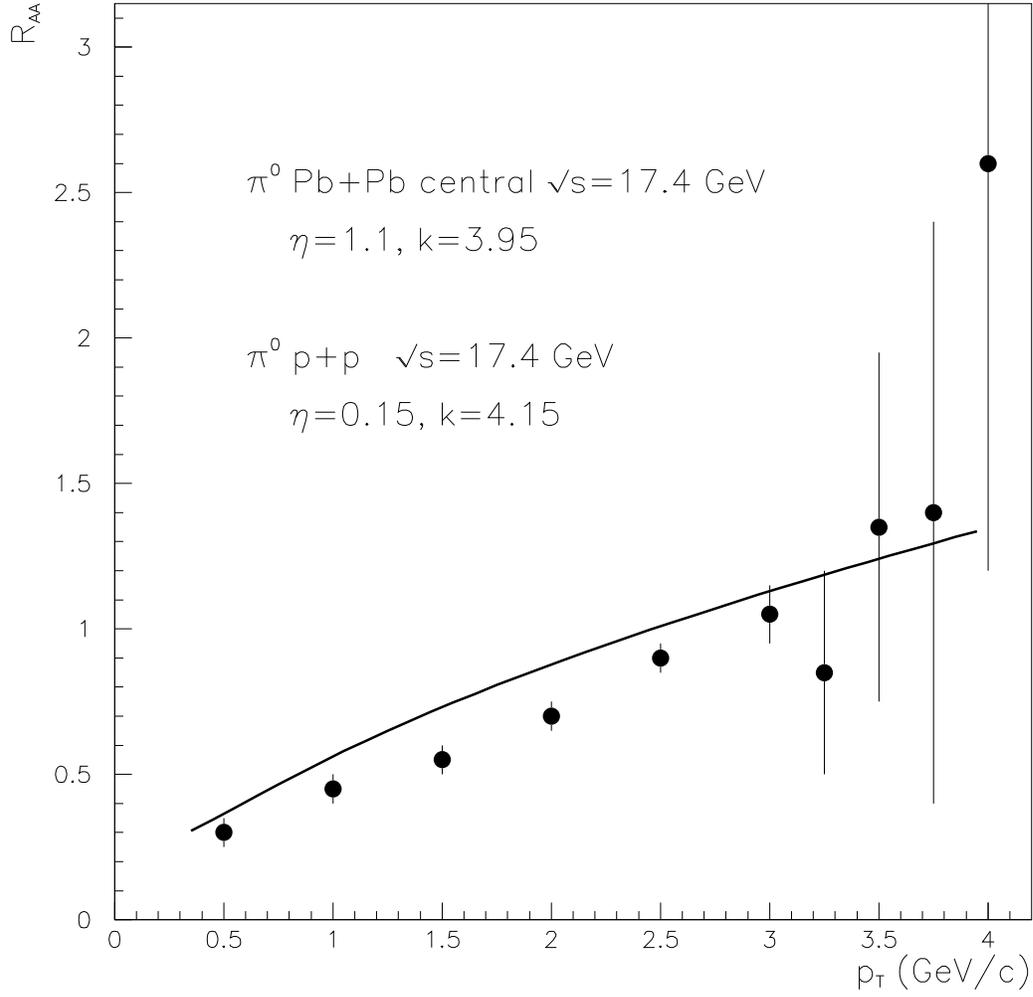}
\vskip 0.5cm
\caption{
Nuclear modification factor for $Pb+Pb$ central collisions at SPS energies.}
\label{figure11}
\end{figure}

\begin{figure}
\centering\leavevmode
\epsfxsize=6in\epsfysize=6in\epsffile{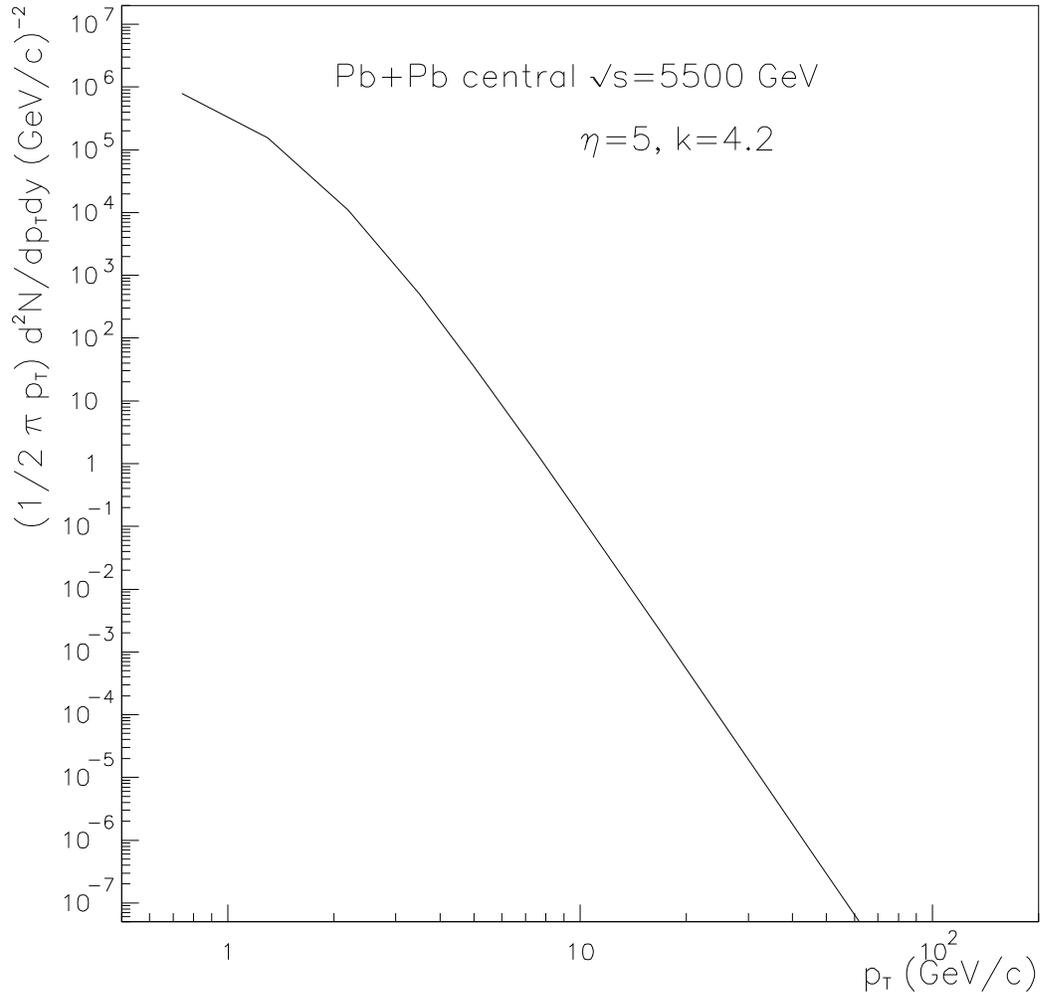}
\vskip 0.5cm
\caption{
Predictions for $Pb+Pb$ central collisions at 5500 GeV.}
\label{figure12}
\end{figure}

\begin{figure}
\centering\leavevmode
\epsfxsize=6in\epsfysize=6in\epsffile{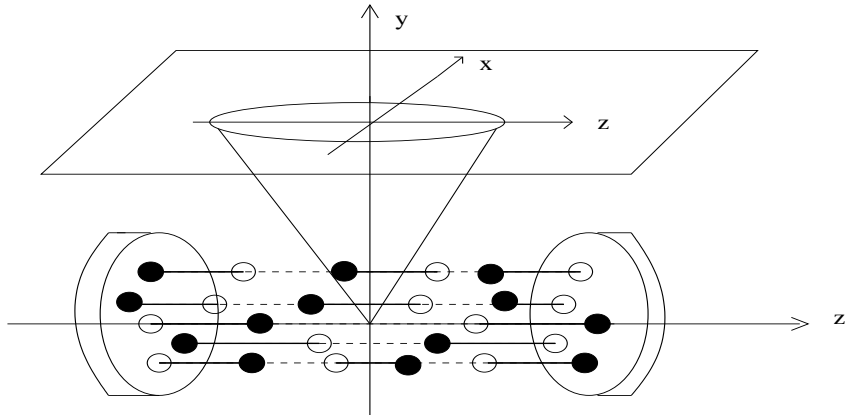}
\vskip 0.5cm
\caption{Formation of strings in a nucleus-nucleus collision.
}
\label{figure13}
\end{figure}

\begin{figure}
\centering\leavevmode
\epsfxsize=6in\epsfysize=6in\epsffile{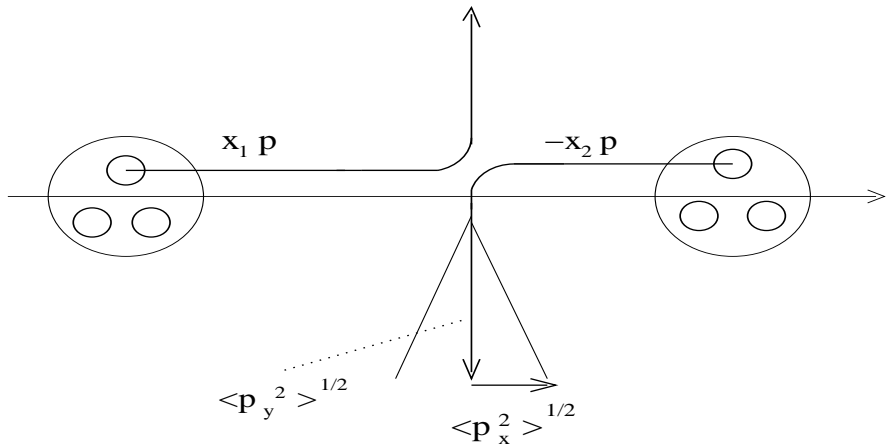}
\vskip 0.5cm
\caption{Hard scattered partons that can interact with the strings. 
}
\label{figure14}
\end{figure}

\end{document}